%% file: ms.tex
\documentclass{article}
\usepackage{ijcai20}

\usepackage{times}

\usepackage{soul}
\usepackage{url}
\usepackage[utf8]{inputenc}
\usepackage[T1]{fontenc}
\usepackage[small]{caption}
\usepackage{amsmath}
\usepackage{booktabs}
\usepackage{pgf}

\input{common/packages}

\input{common/pgfexternalize}

\input{common/commands}

\input{common/math_commands}
\input{common/acronyms}

\setpgfpreamble{

\input{common/packages}
\input{common/pgfexternalize}
\input{common/commands}
\input{common/math_commands}
\input{common/acronyms}\renewcommand{\egoidx}{\ensuremath{i}}
\renewcommand{\otheridx}{\ensuremath{j}}
}

\renewcommand{\egoidx}{\ensuremath{i}}
\renewcommand{\otheridx}{\ensuremath{j}}

\graphicspath{{./pictures/}{./figures/}{./pics/}{./plots}}
\newcommand{\plotpath}{./plots}

\title{	Robust Stochastic Bayesian Games for Behavior Space Coverage}

\author{
	Julian Bernhard$^1$\footnote{Contact Author}\And
	Alois Knoll$^2$\\
	\affiliations
	$^1$fortiss GmbH, An-Institut Technische Universit\"{a}t M\"{u}nchen, Germany\\
	$^2$Chair of Robotics, Artificial Intelligence and Real-time Systems, Technische Universit\"{a}t M\"{u}nchen, Germany\\
	\emails
	bernhard@fortiss.org,
	knoll@mytum.de
}

\begin{document}

\maketitle

\input{./abstract}

\input{./introduction}

\input{./relatedwork}

\input{./preliminaries}

\input{./method}

\input{./experiment}

\input{./conclusion}

\section{Acknowledgement}
This research was funded by the Bavarian Ministry of Economic Affairs, Regional Development and Energy, project Dependable AI.

\footnotesize
\bibliographystyle{named}
\bibliography{bibtex/library.bib}

\end{document}

%% file: common/packages.tex
\usepackage{xargs} 
\usepackage{graphicx}
\usepackage{soul}
\usepackage{pgfgantt}
\usepackage{pdflscape}
\usepackage{lipsum}
\usepackage{pstricks}
\usepackage{import}
\usepackage{xcolor}
\definecolor{midgrey}{RGB}{100, 100, 100}
\definecolor{darkgreen}{RGB}{0, 100, 50}
\definecolor{plum}{RGB}{78, 25, 96}
\usepackage{layouts}
\usepackage{array,booktabs,tabularx}
\usepackage{ragged2e}

\usepackage[colorinlistoftodos,prependcaption,textsize=small]{todonotes}
\usepackage{varwidth}
\usepackage{multirow}
\usepackage{multicol}
\usepackage{siunitx}
\sisetup{load-configurations = abbreviations}

\usepackage{amsmath,amssymb}
\usepackage{mathtools,stackengine}
\usepackage{dsfont}
\DeclareFontFamily{U}{mathc}{}
\DeclareFontShape{U}{mathc}{m}{it}%
{<->s*[1.03] mathc10}{}
\DeclareMathAlphabet{\mathscr}{U}{mathc}{m}{it}

\usepackage{pgf}

\usepackage{algorithm}
\usepackage[noend]{algpseudocode}
\usepackage{scalerel}
\usepackage{stackengine}

\usepackage{url}

\usepackage{caption}

\usepackage[toc, sort=def, acronym, style = long]{glossaries}

%% file: common/pgfexternalize.tex
\makeatletter
\edef\@figdir{_pgfcache}

\catcode`\#=11
\catcode`\|=6
\newcommand{\@basicpgfpreamble}[1]{%
    \unexpanded{%
        \documentclass{standalone}^^J
        \usepackage{pgf}^^J
        \RequirePackage{luatex85}^^J
        \let\oldpgfimage\pgfimage^^J
        \renewcommand{\pgfimage}[2][]{\oldpgfimage[#1]{|1/#2}}^^J
    }%
}
\catcode`\#=6
\catcode`\|=12

\let\@pgfpreamble\@basicpgfpreamble

\newcommand{\setpgfpreamble}[1]{%
    \renewcommand{\@pgfpreamble}[1]{\@basicpgfpreamble{##1}\unexpanded{#1}}
}

\newcounter{@pgfcounter}
\newwrite\@pgfout
\newread\@pgfin

\newcommand{\importpgf}[3][]{%
    \IfFileExists{#2/#3}{}{\errmessage{importpgf: File #2/#3 not found}}%
    \edef\@figfile{\jobname-\the@pgfcounter}%
    \providecommand{\@writetempfile}{}%
    \renewcommand{\@writetempfile}[1]%
    {%
        \immediate\openout\@pgfout=##1%
        \immediate\write\@pgfout{\@pgfpreamble{#2}}%
        \immediate\write\@pgfout{\string\begin{document}}%
        \immediate\openin\@pgfin=#2/#3%
        \begingroup\endlinechar=-1%
            \loop\unless\ifeof\@pgfin%
                \readline\@pgfin to \@fileline%
                \ifx\@fileline\@empty\else%
                    \immediate\write\@pgfout{\@fileline}%
                \fi%
            \repeat%
        \endgroup%
        \immediate\closein\@pgfin%
        \immediate\write\@pgfout{\string\end{document}}%
        \immediate\closeout\@pgfout%
    }%
    \def\@compile%
    {%
        \immediate\write18{lualatex -interaction=batchmode -output-directory="\@figdir" \@figdir/\@figfile.tex}%
    }%
    \IfFileExists{\@figdir/\@figfile.pdf}%
    {%
        \@writetempfile{\@figdir/tmp.tex}%
        \edef\@hashold{\pdfmdfivesum file {\@figdir/\@figfile.tex}}%
        \edef\@hashnew{\pdfmdfivesum file {\@figdir/tmp.tex}}%
        \ifnum\pdfstrcmp{\@hashold}{\@hashnew}=0%
            \relax%
        \else%
            \@writetempfile{\@figdir/\@figfile.tex}%
            \@compile%
        \fi%
    }%
    {%
        \@writetempfile{\@figdir/\@figfile.tex}%
        \@compile%
    }%
    \IfFileExists{\@figdir/\@figfile.pdf}%
    {\includegraphics[#1]{\@figdir/\@figfile.pdf}}%
    {\errmessage{Error during compilation of figure #2/#3}}%
    \stepcounter{@pgfcounter}%
}
\makeatother

%% file: common/commands.tex
\renewcommand{\thefootnote}{\fnsymbol{footnote}}

\newcommandx{\unsure}[2][1=]{\todo[inline,linecolor=red,backgroundcolor=red!25,bordercolor=red,#1]{#2}}
\newcommandx{\info}[2][1=]{\todo[inline,linecolor=OliveGreen,backgroundcolor=OliveGreen!25,bordercolor=OliveGreen,#1]{#2}}
\newcommandx{\improvement}[2][1=]{\todo[inline,linecolor=Plum,backgroundcolor=Plum!25,bordercolor=Plum,#1]{#2}}
\newcommandx{\thiswillnotshow}[2][1=]{\todo[disable,#1]{#2}}
\newcommand{\rework}[1]{%
	\todo[color=yellow!25,inline, caption={}]{\normalsize {#1}}%
}
\newcommand{\idea}[1]{%
	\todo[color=blue!25,inline, caption={}]{\normalsize {#1}}%
}

\newcommand{\important}[1]{%
	\todo[color=red!25,inline, caption={}]{\normalsize {#1}}%
}
\newcommand{\evaluation}[1]{%
	\todo[color=green!25,inline, caption={}]{\normalsize {#1}}%
}
\newcommand{\futurework}[1]{%
	\todo[color=plum!25,inline, caption={}]{\normalsize {#1}}%
}

\graphicspath{{./pictures/}{./figures/}{./../pics/}}

\newcommand{\figurepath}{plots}
\newcommand{\figurepathnew}{../../../Programming/ma_sp/paper_eval/figures/}

\newcommand{\tablepath}{../../../Programming/ma_sp/paper_eval/tables}

\newcolumntype{C}{>{\Centering}X}
\newcolumntype{L}{>{\RaggedRight}X}
\newcolumntype{R}{>{\RaggedLeft}X}

%% file: common/math_commands.tex
\newcommand\equalhat{\mathrel{\stackon[1.5pt]{=}{\stretchto{%
				\scalerel*[\widthof{=}]{\wedge}{\rule{1ex}{3ex}}}{0.5ex}}}}
\DeclareMathOperator{\E}{\mathbb{E}}
\DeclareMathOperator*{\argmax}{argmax}
\DeclareMathOperator*{\argmin}{argmin} 
\DeclareMathOperator*{\infinum}{inf} 
\newcommand{\Rho}{\mathrm{P}}
\newcommand{\Astar}{$\text{A}^* $ }

\renewcommand{\Return}{\gls{Return}}
\newcommand{\Returnmean}{\gls{Returnmean}}
\newcommand{\reward}{\gls{reward}}
\newcommand{\Reward}{\gls{Reward}}
\newcommand{\SThresh}{\gls{SThresh}}

\newcommand{\Real}[1]{\ensuremath{\mathbb{R}^{#1}}} 
\newcommand{\Naturals}[1]{\ensuremath{\mathbb{N}^{#1}}} 
\newcommand{\DistrUniform}{\ensuremath{\mathcal{U}}} 
\newcommand{\BigO}{\ensuremath{\mathcal{O}}} 

%% file: common/acronyms.tex
\newacronym{RL}{RL}{reinforcement rearning}
\glsadd{RL}

\newacronym{cpu}{CPU}{Central Processing Unit}
\glsadd{cpu}

\newacronym{UCB}{UCB}{Upper Confidence Bound}


\newcommand{\est}[1]{\ensuremath{s^{#1}}} 
\newcommand{\EST}[1]{\ensuremath{\mathcal{S}^{#1}_\text{env}}}
\newcommand{\ost}[2]{\ensuremath{o_{#2}^{#1}}} 
\newcommand{\OST}[2]{\ensuremath{\mathcal{O}_{#2}^{#1}}}
\renewcommand{\ast}[2]{\ensuremath{s_{#2}^{#1}}} 
\newcommand{\AST}[2]{\ensuremath{\mathcal{S}_{#2}^{#1}}}
\newcommand{\bst}[3]{\ensuremath{b_{#2#3}^{#1}}} 
\newcommand{\BST}[3]{\ensuremath{\prescript{#3}{}{\mathcal{B}_{#2}^{#1}}}} 
\newcommand{\BSTInt}[3]{\ensuremath{\prescript{#3}{}{B_{#2}^{#1}}}} 
\newcommand{\BSTSize}{\ensuremath{N_B}} 
\newcommand{\hbst}[2]{\ensuremath{\theta_{#2}^{#1}}}
\newcommand{\HBST}[2]{\ensuremath{\Theta_{#2}^{#1}}}
\newcommand{\ist}[2]{\ensuremath{i_{#2}^{#1}}} 
\newcommand{\IST}[2]{\ensuremath{\mathcal{I}_{#2}^{#1}}}
\newcommand{\HIST}[2]{\ensuremath{\prescript{#2}{}{\mathbb{H}}_{I}^{#1}}}

\newcommand{\pst}[1]{\ensuremath{s^{#1}_\text{prior}}} 
\newcommand{\PST}[1]{\ensuremath{\mathcal{S}^{#1}_\text{prior}}} %

\newcommand{\numhyp}{\ensuremath{K}} 
\newcommand{\hypidx}{\ensuremath{k}} 
\newcommand{\hyp}[1]{\ensuremath{\theta_{#1}}} 
\newcommand{\HYP}[1]{\ensuremath{\Theta_{#1}}} 

\newcommand{\sh}[1]{\ensuremath{H^{#1}_\text{s}}}
\newcommand{\SH}{\ensuremath{\mathcal{H}_\text{s}}}

\newcommand{\oh}[1]{\ensuremath{H^{#1}_\text{o}}}
\newcommand{\OH}{\ensuremath{\mathcal{H}_\text{o}}}

\renewcommand{\a}[3]{\ensuremath{a_{#2#3}^{#1}}}
\newcommand{\A}[2]{\ensuremath{A_{#2}^{#1}}}

\newcommand{\p}[1]{\ensuremath{\pi_{#1}}}
\newcommand{\ptrue}[1]{\ensuremath{\pi_{#1}}}
\newcommand{\phyp}{\ensuremath{\pi^{*}}}

\newcommand{\TEnv}{\ensuremath{T_\text{env}}}
\newcommand{\TPrior}{\ensuremath{T_\text{prior}}}
\newcommand{\TPriorSpace}{\ensuremath{\mathcal{T}_\text{prior}}}
\newcommand{\THypo}{\ensuremath{T_\text{hypo}}}

\newcommand{\PEnv}{\ensuremath{\mathbb{P}_\text{env}}}
\newcommand{\PPrior}{\ensuremath{\mathbb{P}_\text{prior}}}

\newcommand{\Smeas}{\ensuremath{f}}
\newcommand{\GRisklevel}{\ensuremath{\beta}}
\newcommand{\SRisklevel}{\ensuremath{\beta_\text{scenario}}}
\newcommand{\SRisk}{\ensuremath{\rho_\text{scenario}}}
\newcommand{\DRisk}{\ensuremath{\rho_\text{decision}}}
\newcommand{\HRisk}{\ensuremath{\rho_\text{history}}}

\newcommand{\Runtime}{\ensuremath{\mathsf{T}}}

\newcommand{\TScen}{\ensuremath{\mathsf{T}_\text{sc}}}

\newcommand{\egoidx}{\ifmmode \text{ego}\else ego\fi}
\newcommand{\otheridx}{\ifmmode \text{other}\else other\fi}
\newcommand{\egonumact}[2]{\ensuremath{k_\egoidx}}
\newcommand{\numotheragents}{\ensuremath{N_\otheridx}}

\newcommand{\discount}{\ensuremath{\gamma}}

\newcommand{\desiredgap}[2]{\ensuremath{d^{#1}_{#2}}}
\newcommand{\desiredgapleft}[1]{\ensuremath{d_{l,#1}}}
\newcommand{\desiredgapright}[1]{\ensuremath{d_{r,#1}}}

\newcommand{\idmdesiredvelocity}{\ensuremath{v_\text{desired}}}
\newcommand{\idmminimumspacing}{\ensuremath{s_\text{min}}}
\newcommand{\idmdesiredheadway}{\ensuremath{T_\text{desired}}}
\newcommand{\idmaccfactor}{\ensuremath{\dot{v}_\text{factor}}}
\newcommand{\idmaccmin}{\ensuremath{\dot{v}_\text{min}}}
\newcommand{\idmaccmax}{\ensuremath{\dot{v}_\text{max}}}
\newcommand{\idmcomftbrake}{\ensuremath{\dot{v}_\text{comft}}}

\newcommand{\acccoolness}{\ensuremath{C_\text{coolness}}}

\newglossaryentry{Return}{%
	name=\ensuremath{R},
	description={Random Return Variable distributed according to return distribution $R\sim Z$}
}

\newglossaryentry{Returnmean}{%
	name=\ensuremath{\overline{R}},
	description={Mean of return $\overline{R} = \E_{R\sim Z}[R]$}
}

\newglossaryentry{reward}{%
	name=\ensuremath{\mathscr{r}},
	description={single reward value given for transition from $s$ to $s^\prime$ executing action $a$}
}

\newglossaryentry{Reward}{%
	name=\ensuremath{\mathcal{R}},
	description={Reward distribution for transition from $s$ to $s^\prime$ executing action $a$}
}

\newglossaryentry{SThresh}{%
	name=\ensuremath{\lambda_\text{safety}~},
	description={Defines the amount of safety the decision making algorithm shall achieve}
}

\newglossaryentry{EnvStateT}{%
	name=\est{t},
	description={The environment state at time $t$, $\est{t}=(\ast{t}{0}, \ast{t}{1},\ldots,\ast{t}{n})$}
}

\newglossaryentry{EnvStateSpace}{%
	name=\EST{},
	description={The space of environment states}
}

\newglossaryentry{PriorEnvStateT}{%
	name=\pst{t},
	description={The prior environment state at time $t$, $\pst{t}=(\ast{t}{0}, \ast{t}{1},\ldots,\ast{t}{n})$}
}

\newglossaryentry{PriorEnvStateSpace}{%
	name=\PST{},
	description={The space of prior environment states shall be defined as a superset of the environment state space $\PST{} \supseteq \EST{}$. }
}

\newglossaryentry{AgentStateT}{%
	name=\ast{t}{i},
	description={The full state of agent $i$ at time $t$, $\ast{t}{i} =(\ost{t}{i},\ist{t}{i}, \bst{t}{i} )$}
}

\newglossaryentry{AgentStateSpace}{%
	name=\AST{}{i},
	description={The state space of agent $i$, $\AST{}{i} = \OST{}{i} \times \IST{}{} \times \BST{}{i}{}$ }
}

\newglossaryentry{ObservableAgentStateT}{%
	name=\ost{t}{i},
	description={The observable state of agent $i$ at time $t$, in our case the physical state}
}

\newglossaryentry{ObservableAgentStateSpace}{%
	name=\OST{}{i},
	description={The space of observable states of agent $i$}
}

\newglossaryentry{ObservableStateT}{%
	name=\ost{t}{},
	description={The observable state of the environment at time $t$, in our case the physical states of all observable agents}
}

\newglossaryentry{BehaviorAgentStateT}{%
	name=\bst{t}{i}{},
	description={The behavior state of agent $i$ at time $t$  }
}

\newglossaryentry{BehaviorStateT}{%
	name=\bst{t}{}{},
	description={The environment behavior state of all observable agents at time $t$,$\bst{t}{}{}=(\bst{t}{0}{},\bst{t}{1}{},\ldots,\bst{t}{n}{})$   }
}

\newglossaryentry{BehaviorStateSpace}{%
	name=\BST{}{i}{},
	description={The space of behavior states}
}

\newglossaryentry{IntentionAgentStateT}{%
	name=\ist{t}{i},
	description={The intention state of agent $i$ at time $t$ }
}

\newglossaryentry{IntentionStateT}{%
	name=\ist{t}{},
	description={The environment intention state of all observable agents at time $t$, $\ist{t}{}=(\ist{t}{0},\ist{t}{1},\ldots,\ist{t}{n})$ }
}

\newglossaryentry{IntentionStateSpace}{%
	name=\IST{}{},
	description={The space of intentions}
}

\newglossaryentry{StateHistoryT}{%
	name=\sh{t},
	description={The state action history up to time $t$ starting from the initial observation state \ost{0}, $H^t=(\est{0},\a{0}{}{},\est{1},\a{1}{}{},\ldots,\est{t})$}
}

\newglossaryentry{StateHistorySpace}{%
	name=\SH,
	description={The space of state action histories.}
}

\newglossaryentry{ObservationHistoryT}{%
	name=\oh{t},
	description={The observation action history up to time $t$ starting from the initial observation state \ost{0}, $H^t=(\ost{0},\a{0}{}{},\ost{1},\a{1}{}{},\ldots,\ost{t})$}
}

\newglossaryentry{ObservationHistorySpace}{%
	name=\OH,
	description={The space of observation action histories }
}

\newglossaryentry{AgentAction}{%
	name=\a{t}{i}{k},
	description={The $k$th action of agent $i$ at time $t$}
}

\newglossaryentry{JointAction}{%
	name=\a{t}{}{},
	description={The joint action $\a{t}{}{}=(\a{t}{1}{},\a{t}{2}{},\ldots,\a{t}{n}{})$ selected by the agents at time $t$}
}

\newglossaryentry{ActionSet}{%
	name=\A{t}{i},
	description={The set of possible actions of agent $i$ at time $t$, $\A{t}{i}=(\a{t}{i}{1},\a{t}{i}{2},\ldots,\a{t}{i}{k_i})$ where $k_i^t$ is the number of actions of agent $i$ at time $t$ }
}

\newglossaryentry{AgentPolicy}{%
	name=\p{i},
	description={The stochastic policy of agent $i$ mapping action observations histories \oh{t} to action \a{t}{}{} with $\p{i} : \OH \times \A{t}{i} \rightarrow [0,1] $ }
}

\newglossaryentry{WorldTransition}{%
	name=\TEnv,
	description={The stochastic environment transition function maps the ego-agents action \a{t}{0}{} to the next state: $\TEnv: \A{}{0} \times \OH{} \times \EST{} \rightarrow [0,1]$     }
}

\newglossaryentry{EnvironmentDistribution}{%
	name=\PEnv,
	description={The environment distribution models what environment states \est{0} and transition functions \TEnv are encountered initially at the start of a scenario: $\est{0}, \TEnv \sim \PEnv$    }
}

\newglossaryentry{PriorEnvironmentDistribution}{%
	name=\PPrior,
	description={The prior environment distribution models what prior environment states \pst{0} and transition functions \TPrior are encountered initially at the start of a scenario: $\pst{0},\TPrior \sim \PPrior$    }
}


\newacronym[plural=sbg,firstplural=Stochastic Bayesian Games (SBGs)]{sbg}{SBG}{Stochastic Bayesian Game}
\newacronym{mdp}{MDP}{Markov Decision Process}
\newacronym{rmdp}{RMDP}{Robust Markov Decision Process}
\newacronym[plural=rsbg,firstplural=Robust Stochastic Bayesian Games (RSBGs)]{rsbg}{RSBG}{Robust Stochastic Bayesian Game}
\newacronym{hba}{HBA}{Harsanyi Bellman Ad Hoc}
\newacronym{bamdp}{BAMDP}{Bayesian-adaptive \gls{mdp}}

\newacronym{mcts}{MCTS}{Monte Carlo Tree Search}

%% file: abstract.tex
\begin{abstract}
	A key challenge in multi-agent systems is the design of intelligent agents solving real-world tasks in close interaction with other agents (e.g. humans), thereby being confronted with a variety of behavioral variations and limited knowledge about the true behaviors of observed agents.
	The practicability of existing works addressing this challenge is being limited due to using finite sets of hypothesis for behavior prediction, the lack of a hypothesis design process ensuring coverage over all behavioral variations and sample-inefficiency when modeling continuous behavioral variations.  In this work, we present an approach to this challenge based on a new framework of \emph{\glspl{rsbg}}. An RSBG defines hypothesis sets by partitioning the physically feasible, continuous behavior space of the other agents. It combines the optimality criteria of the \gls{rmdp} and the \gls{sbg} to exponentially reduce the sample complexity for planning with hypothesis sets defined over continuous behavior spaces. Our approach outperforms the baseline algorithms in two experiments modeling time-varying intents and large multidimensional behavior spaces, while achieving the same performance as a planner with knowledge of the true behaviors of other agents.
    \end{abstract}

%% file: introduction.tex
\section{Introduction}
	
    Autonomous agents must be able to solve complex, real-world tasks in close interaction with humans. In many tasks there remain only a few seconds of observations for the agent to adapt its plan to the behavior of the participating humans. Important examples include intersection crossing of an autonomous vehicle or robot navigation through dense pedestrian areas. Among the variety of options to model this partially-cooperative multi-agent problem \cite{albrecht_autonomous_2018}, the \glsfirst{sbg} \cite{albrecht_game-theoretic_2013} is particularly qualified: It uses a predefined finite set of behavior hypothesis for the other agents to adapt to the of observed behavior of other agents during the interaction process  \cite{albrecht_reasoning_2017,stone_ad_2010}. Each hypothesis is commonly defined as probability distribution mapping observation histories to actions. The hypothesis set can either be learned from data of interaction histories \cite{barrett_teamwork_2013,barrett_cooperating_2015} or defined by domain experts \cite{barrett_empirical_2011,ravula_ad_2019}. However, the following shortcomings exist with the  \gls{sbg} to deal with the prescribed problem: 1)~It is unclear how to define the set of hypothesis to cover every \emph{physically} feasible human behavior. Data-based methods do often neglect edge-cases in human behavior as they may not be observed during the data recording process. Domain experts do not have any method at hand to design hypothesis sets covering the complete human behavior space. 2)~The  \gls{sbg} is defined for a limited, finite number of hypothesis. However, a finite hypothesis set is unable to express the subtle, \emph{continuous} variations inherent to human behavior.
    
	To clarify the two shortcomings, we exemplarily define the hypothesis set for an autonomous vehicle having to cross an intersection. A domain expert could assume that a human has two intentions in this task with respect to other vehicles, "give way" or "take way". It would directly map these to a set of two corresponding behavioral hypothesis. This definition describes \emph{what} may happen in the intersection leaving unclear how to further partition the hypothesis set to model \emph{how} this \emph{physically} happens. For instance, giving way can be realized at various distances to the other vehicle. One can suggest to learn a mapping from intents to physical realizations. However, edge-cases in behavior, e.g emergency braking, are rarely recorded and may thus not be adequately represented in a learned hypothesis set. On the other hand, since human intents are not \emph{physically} measurable, this impedes a definition of a ground truth label set. Learned mappings from intent models to physical behavior might thus be incorrect to a certain extent.  

    The goal of this work is to overcome these shortcomings. We present a design process for hypothesis sets achieving coverage over the \emph{physically} feasible, \emph{continuous} behavior space of other agents.  The behavior space is defined such that it comprises all physically feasible behavioral variations and can straightforwardly be defined by a domain expert. To reduce the sample complexity when planning with hypothesis sets defined over a continuous space, we formulate the  \emph{\glsfirst{rsbg}}. It integrates the worst-case optimality criterion of the  \gls{rmdp} \cite{nilim_robust_2005-1} into the Harsanyi-Bellman optimality equation \cite{albrecht_utilising_2015} of the \gls{sbg}. We present a variant of \gls{mcts} to solve the  \gls{rsbg}. Finally, in an intersection crossing task with broad behavioral variations of other agents and a lane changing task with a large multidimensional behavior space, we find that our approach outperforms the  \gls{sbg} in the average number of successful trials and achieves the same performance as a planning algorithm with knowledge of the true behavior of other agents.

%% file: relatedwork.tex
\section{Related Work}
In this section, we discuss methods of hypothesis definition for the \gls{sbg}. Next, we present the  \gls{rmdp} and its link to our research.
\subsection{Hypothesizing Behaviors}   
Previous works frequently use small hypothesis sets in simpler domains defined by domain experts  \cite{stone_ad_2010,albrecht_game-theoretic_2013}. Discrete sets of behavior hypothesis are also frequently employed in robotics with intention-based agent models \cite{bai_intention-aware_2015,tamura_development_2012,sadigh_information_2016}. 
As previously discussed, we consider discrete hypothesis sets as inadequate to cover all behavioral variations emerging in real-world tasks.

Integrating continuity into behavior hypothesis can be broadly categorized into approaches using a parameterized set of hypothesis or approaches learning a hypothesis set on the fly during task completion. Methods in the former category either  build a hypothesis set by sampling hypothesis out of a parameterized hypothesis space \cite{southey_bayes_2005}  or adapt online the parameters of a predefined set of hypothesis \cite{hindriks_opponent_2008,albrecht_reasoning_2017}. However, such methods only consider a single parameter set for each hypothesis and do not model types which \emph{cover} a certain part of the parameter space. For instance,  instead of modeling the preferred distance of a specific agent to other vehicles in an intersection as fixed, single parameter, it should be defined as varying slightly over time to express the subtle continuous behavioral variations in human behavior. 
With Q-learning \cite{barrett_cooperating_2015} or decision trees \cite{barrett_empirical_2011} the hypothesis set can be adapted on the fly avoiding the definition of a continuous hypothesis model.
However, online adaptation of the hypothesis set is impractical when the task is characterized by short interaction times as considered in this work. 
In addition to the mentioned shortcomings, all of the above works do not specify a \emph{hypothesis design process} to achieve coverage over all possible agent types. In our work, a hypothesis set partitions a behavioral space, defined by a domain expert. This process may be a potential solution to this open question.

\subsection{Robustness-Based Optimality}
The robustness of a plan or policy to \emph{continuous} modeling errors has long been studied in the control and reinforcement learning community \cite{bagnell_solving_2001,nilim_robust_2005-1,li_robust_2019,lim_reinforcement_2013}. The \glsfirst{rmdp} framework searches for a solution which is optimal under the worst-case parameter realizations of a (possibly continuous \cite{tamar_scaling_2014}) set of parameters of the transition function, denoted uncertainty set. The main challenge with the robustness criterion is finding an uncertainty set which avoids overly conservative policies \cite{derman_bayesian_2019,petrik_beyond_2019}.

Combinations of robust optimization and Bayesian decision making have been investigated in reinforcement learning \cite{derman_bayesian_2019} and game theory \cite{aghassi_robust_2006}. The latter approach, denoted Robust game theory, applies the worst-case operation \emph{over the type space} to omit dependency on posterior type-beliefs in the expected value calculation.  In contrast to their work, we split the continuous parameter space into multiple uncertainty sets and apply the worst-case operation \emph{over the parameter space} of each type. The outcomes are then weighted with the posterior belief of each type. This method allows to cover a continuous parameter space and to control the conservativeness of the policy via the number of defined types.

%% file: preliminaries.tex
\section{Preliminaries}

We propose a mathematical definition of behavior spaces and present background on the \gls{sbg} and \gls{rmdp}.

\subsection{Behavior Spaces} \label{sec:environment_model}

We consider a multi-agent environment with $N$ interacting agents. The process starts at time $t=0$. At time step $t$, each agent $j$ observes the joint environment state $\ost{t}{}=(\ost{t}{1},\ost{t}{2},\ldots,\ost{t}{N})$ and chooses an action \a{t}{j}{} from a continuous action space \A{}{j}{}. The environment state \ost{t}{} describes the current \emph{physical} properties, e.g. position, velocity, etc. Based on the agents' joint action $\a{t}{}{}\in \A{}{}{}=\times \A{}{j}{}$ with joint action space \A{}{}{} the environment transitions to the next state, $\ost{t+1}{}$. We leave the precise transition model open. This process continues until some terminal criterion is satisfied.

An agent chooses an action \a{t}{j}{} according to its policy $\a{t}{j}{}\sim \ptrue{\otheridx}(\a{t}{j}{}|\oh{t}{}, \ist{t}{j})$. The policy depends on the observation action history up to time $t$, $\oh{t}{}=(\ost{0}{},\a{0}{}{},\ost{1}{},\a{1}{}{},\ldots,\ost{t}{})$ and a time-dependent intention state \ist{t}{j}. The intention state may encode long- or short-term abstract goals or a more precise plan. We leave the exact model and dynamics of the intention state open. 

\begin{figure}[t]
	\def\svgwidth{\columnwidth}
	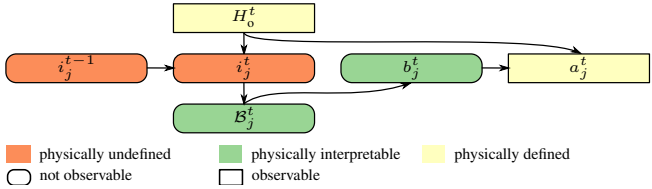 
	\caption{Causal diagram to model the conditional dependence of intentions, behavior space and state, and actions for other agents \otheridx. Behavior spaces \BST{t}{\otheridx}{} are affected by intent states $\ist{t}{\otheridx}$ and span a range of possible behavior states $\bst{t}{\otheridx}{}$ upon which the other agent's policy depends.}
	\vspace{-4mm}
	\label{fig:causal_model_behavior_space}
\end{figure}

We control a single agent, $\egoidx$, which reasons about the behavior of the other agents $\otheridx$. We assume that $i$ knows the action space and can observe past actions of the other agents. The true policy \ptrue{\otheridx} and any intent information of the other agent are \emph{unknown} to \egoidx. However, we assume for a specific task there exists a single hypothetical policy \begin{equation}
\phyp: \OH \times  \BST{t}{j}{}  \rightarrow \A{}{\otheridx}
\end{equation} with $\bst{t}{\otheridx}{}\in\BST{}{j}{}$ being agent's \otheridx~ behavior state at time $t$, $\BST{t}{j}{} \subset \Real{\BSTSize}$ its behavior space of dimension \BSTSize~ and \OH~ the space of all action observation histories.  The hypothetical policy is defined such that a behavior state \bst{t}{\otheridx}{} is a \emph{physically interpretable} quantity describing \otheridx's behavior at \emph{the time point of interaction}. Agent \otheridx~covers its behavior space \BST{}{j}{} by sampling its behavior state \bst{t}{\otheridx}{} uniformly from \BST{}{j}{} in each time step, $\bst{t}{\otheridx}{}\sim\DistrUniform(\BST{}{j}{})$ before choosing an action according to $\phyp$. In our model, solely \BST{}{j}{} depends on the intention state, whereas the policy is independent. 
The causal diagram in Fig.~\ref{fig:causal_model_behavior_space} illustrates the relations between the random variables in our model\footnote{Causal models define an interventional type of conditional distribution instead of the observational variant \cite{pearl_causality_2000}. We are interested in $\phyp(a|\oh{}{}, do(\bst{t}{\otheridx}{}))$ and not in $\phyp(\a{t}{\otheridx}{}|\oh{}{}, \bst{t}{\otheridx}{})$. For the latter definition the joint distribution $p(\a{t}{\otheridx}{}, \oh{}{}, \bst{t}{\otheridx}, \ist{t}{\otheridx})$ must exist which is not the case as intents cannot be measured.}.

The other agents' behavior spaces $\BST{t}{j}{}$ and their current behavior state \bst{t}{\otheridx}{} are not observable. However, using the property of physical interpretability of \bst{t}{\otheridx}{}, an expert can define a full behavior space \BST{}{}{}, comprising the individual behavior spaces $\BST{}{j}{}$ ($\BST{}{j}{} \subset \BST{}{}{}$), by looking at the physically realistic situations. For instance, it is straightforward to define the physical boundaries of a behavior state modeling the desired gap between agent \otheridx~ and \egoidx~ at the time point of crossing an intersection with the one-dimensional behavior space $\BST{}{}{}=\{\bst{}{}{}| \bst{}{}{}\in [-d_\text{max}, d_\text{max}]\}$ where $d_\text{max}$ is the maximum sensor range. 

In the remainder of this paper, we design a decision model enabling sample-efficient planning for agent \egoidx~ based on the hypothetical policy \phyp~ and hypothesis sets defined over the full behavior space \BST{}{}{}.

\subsection{Harsanyi-Bellman Ad Hoc Algorithm}

The type-based approach \cite{albrecht_belief_2016} uses a predefined set of behavior types  $\hbst{}{\hypidx} \in \HBST{}{}$ and hypothetical behavior policies $\a{t}{j}{} \sim \p{\hbst{\hypidx}{}}(\a{t}{j}{}|\oh{t}{})$ for the other agents \otheridx. Given the action-observation history of an agent one can track a posterior belief $\text{Pr}(\hbst{\hypidx}{\otheridx}|\oh{t}{}) \sim L(\oh{t}|\hbst{\hypidx}{\otheridx})P(\hbst{}{\hypidx})$ over hypothesized types over time with $P(\hbst{}{\hypidx})$ being the prior of a type. Depending on the calculation of the likelihood $L(\cdot)$, one obtains either a product or sum posterior.

In the remainder of this paper, an index~$-i$ denotes all agents except $i$, giving for the joint action $\a{}{}{}{=}\a{}{i,-\egoidx}{}$ and the joint type space of other agents $\HBST{}{-\egoidx}{=}\times_{j=1}^{N, j\neq i}\HBST{}{}$.
The \gls{hba} algorithm \cite{albrecht_game-theoretic_2013} plans an optimal action for agent \egoidx~ according to the optimality criterion $\a{t}{\egoidx}{} \sim \argmax_{\a{}{\egoidx}{}} E_{\ost{}{}}^{\a{}{\egoidx}{}}(\oh{t}{})$, where $E^{\a{}{\egoidx}{}}_{\ost{}{}}(\oh{\prime}{})=$
\begin{equation}  \label{eq:HBA_expectation}
 \sum_{\hbst{}{-\egoidx}\in \HBST{}{-\egoidx}}\text{Pr}(\hbst{}{-\egoidx}|\oh{t}{})\sum_{\a{}{-\egoidx}{}\in\A{}{-\egoidx}{}}Q^{\a{}{i,-\egoidx}{}}_{\ost{}{}}(\oh{\prime}{})\prod_{\otheridx\neq \egoidx} \p{\hbst{}{j}}(\oh{\prime}{}, \a{}{j}{})
\end{equation}
is the expected cumulative reward for agent $i$ taking action \a{}{i}{} in state \ost{}{} and history \oh{\prime}{}. The Bellman part of HBA\footnote{As we consider deterministic joint transition functions, we can neglect the expectation over potential subsequent states $s^\prime$.} is $Q^{\a{}{}{}}_{\ost{}{}}(\oh{\prime}{})=$
\begin{equation}
r(\ost{}{},\a{}{}{}) + \discount \max_{\a{}{i}{}\in \A{}{\egoidx}} E^{\a{}{i}{}}_{\ost{\prime}{}}(\left\langle\oh{\prime}{}, \a{}{}{}, \ost{\prime}{} \ \right\rangle)
\label{eq:HBA_bellman}
\end{equation}
and defines the expected cumulative future reward of agent \egoidx~ when joint action \a{}{}{} is executed in observation state \ost{}{} after history \oh{\prime}{}. Future rewards are discounted by $\discount$. 
\glsfirst{mcts} can be used to find approximate solutions to this equation \cite{barrett_teamwork_2013}.

\subsection{Adversarial Reasoning}
 A \glsfirst{rmdp} models uncertainty about the parameters of the transition function $p$ in an \gls{mdp}  \cite{nilim_robust_2005-1}. Its optimality criterion \mbox{$\a{t}{i}{} \sim \argmax_{\a{}{i}{}} Q^{\a{}{}{}}_{\ost{}{}}$} can be seen as two-agent stochastic game where an adversary tries to minimize the expected cumulative future reward of the controlled agent by picking the transition function $p$ inducing the worst-case outcome. The robust Bellman equation \cite{tamar_scaling_2014} is defined as
 \begin{equation*}
  Q^{\a{}{}{}}_{\ost{}{}} = r(\ost{}{},\a{}{}{}) + \discount \max_{\a{}{i}{}} 
   \inf_{p\in \mathcal{P}} \E^p [Q^{\a{}{}{}}_{\ost{\prime}{}}|\ost{}{}, \a{}{}{}].
 \end{equation*}
In the multi-agent case, with limited knowledge about the policies of other agents, we apply the worst-case assumption over other agents' actions to get the robust Bellman equation  
 \begin{equation} \label{eq:robust_multi_agent}
 Q^{\a{}{}{}}_{\ost{}{}} = r(\ost{}{},\a{}{}{}) + \discount \max_{\a{}{i}{}\in \A{}{\egoidx}} 
 \min_{\a{}{-\egoidx}{}\in\A{}{-\egoidx}{}} Q^{\a{}{\egoidx, -\egoidx}{}}_{\ost{\prime}{}}
 \end{equation} 
with minimax learning objective \cite{li_robust_2019}.

%% file: pics/causal_diagram_behavior_space.pdf_tex
\begingroup%
  \makeatletter%
  \providecommand\color[2][]{%
    \errmessage{(Inkscape) Color is used for the text in Inkscape, but the package 'color.sty' is not loaded}%
    \renewcommand\color[2][]{}%
  }%
  \providecommand\transparent[1]{%
    \errmessage{(Inkscape) Transparency is used (non-zero) for the text in Inkscape, but the package 'transparent.sty' is not loaded}%
    \renewcommand\transparent[1]{}%
  }%
  \providecommand\rotatebox[2]{#2}%
  \ifx\svgwidth\undefined%
    \setlength{\unitlength}{1755.3674797bp}%
    \ifx\svgscale\undefined%
      \relax%
    \else%
      \setlength{\unitlength}{\unitlength * \real{\svgscale}}%
    \fi%
  \else%
    \setlength{\unitlength}{\svgwidth}%
  \fi%
  \global\let\svgwidth\undefined%
  \global\let\svgscale\undefined%
  \makeatother%
  \begin{picture}(1,0.28206569)%
    \put(0.37054963,0.08119963){\color[rgb]{0,0,0}\makebox(0,0)[b]{\smash{}}}%
    \put(0,0){\includegraphics[width=\unitlength,page=1]{causal_diagram_behavior_space.pdf}}%
    \put(0.37090707,0.09807433){\color[rgb]{0,0,0}\makebox(0,0)[b]{\smash{\tiny  $\BST{t}{\otheridx}{}$}}}%
    \put(0.37090707,0.25371941){\color[rgb]{0,0,0}\makebox(0,0)[b]{\smash{\tiny $\oh{t}$}}}%
    \put(0.11109173,0.17589687){\color[rgb]{0,0,0}\makebox(0,0)[b]{\smash{\tiny $\ist{t-1}{\otheridx}$}}}%
    \put(0.63072242,0.17589687){\color[rgb]{0,0,0}\makebox(0,0)[b]{\smash{\tiny $\bst{t}{\otheridx}{}$}}}%
    \put(0,0){\includegraphics[width=\unitlength,page=2]{causal_diagram_behavior_space.pdf}}%
    \put(0.37090706,0.17589687){\color[rgb]{0,0,0}\makebox(0,0)[b]{\smash{\tiny $\ist{t}{\otheridx}$}}}%
    \put(0,0){\includegraphics[width=\unitlength,page=3]{causal_diagram_behavior_space.pdf}}%
    \put(0.89033526,0.17589687){\color[rgb]{0,0,0}\makebox(0,0)[b]{\smash{\tiny $\a{t}{\otheridx}{}$}}}%
    \put(0,0){\includegraphics[width=\unitlength,page=4]{causal_diagram_behavior_space.pdf}}%
    \put(0.05289903,0.04095507){\color[rgb]{0,0,0}\makebox(0,0)[lb]{\smash{\tiny physically undefined}}}%
    \put(0.38164937,0.04095507){\color[rgb]{0,0,0}\makebox(0,0)[lb]{\smash{\tiny physically interpretable}}}%
    \put(0.05343883,0.00688756){\color[rgb]{0,0,0}\makebox(0,0)[lb]{\smash{\tiny not observable}}}%
    \put(0,0){\includegraphics[width=\unitlength,page=5]{causal_diagram_behavior_space.pdf}}%
    \put(0.3822256,0.00688756){\color[rgb]{0,0,0}\makebox(0,0)[lb]{\smash{\tiny observable}}}%
    \put(0,0){\includegraphics[width=\unitlength,page=6]{causal_diagram_behavior_space.pdf}}%
    \put(0.6965433,0.04095507){\color[rgb]{0,0,0}\makebox(0,0)[lb]{\smash{\tiny physically defined}}}%
  \end{picture}%
\endgroup%

%% file: method.tex
\section{Method}
In this section, we first present a design process for hypothesis sets to achieve behavior space coverage based on our environment model from sec.~\ref{sec:environment_model}. Next, we discuss the \gls{rsbg} and our variant of \gls{mcts} to enable sample-efficient planning with our hypothesis definition.

\subsection{Hypothesis Sets for Behavior Space Coverage} \label{sec:hypothesis_sets_coverage}
The standard type-based method tries to define each hypothetical type \hbst{\hypidx}{} such that it can closely match a \emph{single} unknown policy  \ptrue{\otheridx} of another agent \otheridx. In contrast, we define a collection of hypothesis each covering a certain part of the continuous behavior space \BST{}{}{}. Thus, multiple hypothesis equally participate in representing an unknown policy \ptrue{\otheridx}. 

Specifically, we define a partition of the full behavior space $\BST{}{}{}=\BST{1}{}{} \cup \BST{2}{}{} \cup \ldots \cup \BST{\numhyp}{}{}, \, \forall t\neq k \BST{t}{}{} \cap \BST{k}{}{} = \emptyset$ to form \numhyp~hypothesis
$\p{\hbst{\hypidx}{}}: \OH \times  \A{\hypidx}{} \rightarrow [0,1], \,k\in \{1,\ldots,\numhyp\}$. To define the hypothesis, we need a probability distribution over actions. We define this distribution in terms of the hypothetical policy $\phyp$ and the part \BST{\hypidx}{}{}. In sec.~\ref{sec:environment_model}, we define that an agent covers its behavior space \BST{}{j}{} by sampling a behavior state \bst{t}{j}{} from a uniform distribution in each time step $t$. Therefore, we can use a uniform density over behavior states $f(\bst{}{}{})=\tfrac{1}{||\BST{\hypidx}{}{}||_V}$ to define the hypothesis set (with $||\cdot||_V$ measuring the volume of a space), and obtain 
\begin{equation}
\p{\hbst{\hypidx}{}}(\a{t}{\otheridx}{}| \oh{t}{}) = \text{Pr}(\{\bst{}{}{}\,|\,\forall \bst{}{}{}\in\BST{\hypidx}{}{},\,\phyp(\bst{}{}{},\oh{t}{}) = \a{t}{\otheridx}{} \} )
\end{equation}
with action space $\A{\hypidx}{}=\{\a{}{}{}| \,\forall \bst{}{}{}\in\BST{\hypidx}{}{},\,\phyp(\bst{}{}{},\oh{t}{}) = \a{}{}{}\}$. The action space \A{\hypidx}{} becomes \emph{continuous}, since different behavior states typically imply different actions and we have $|\A{\hypidx}{}|\approx|\BST{\hypidx}{}{}|$ where $|\cdot|$ is an abstract measure of how many samples sufficiently represent the underlying continuous space.

\subsection{Robust Stochastic Bayesian Games}
Approximating a solution to eq.~\ref{eq:HBA_expectation} with \gls{mcts} is computationally demanding for a continuous space of joint actions $\A{}{-\egoidx}=\times_{k\in \hbst{}{-\egoidx}}\A{\hypidx}{}$.
To get further insight into the problem, we calculate the sample complexity of eq.~\ref{eq:HBA_expectation} for our hypothesis definition: Using equal-sized partitions of \BST{}{}{}, we get $|\A{\hypidx}{}|\approx|\BST{}{}{}|/\numhyp$ and obtain for the size of the joint action space  $|\A{}{-i}|{=}\prod_{\otheridx=1}^{N^\prime}|\A{\hypidx}{}|{\approx}(|\BST{}{}{}|/\numhyp)^{N^\prime}$ with $N^\prime {=} N{-}1$ being the number of other agents.   
Sampling over a joint action space \A{}{-\egoidx} occurs for all combinations of types $\hbst{}{-\egoidx}{\in} \HBST{}{-\egoidx}$ with $|\HBST{}{-\egoidx}|{=} K^{N^\prime}$ whereas \hbst{}{-\egoidx} is sampled once in each iteration \cite{barrett_teamwork_2013}.  Since different joint actions can occur at each prediction time step $t$, this introduces an additional exponent and we get a sample complexity
$\BigO(|\HBST{}{-\egoidx}|\cdot|\A{}{-\egoidx}|^t) = \BigO_\text{SBG}(|\BST{}{}{}|^{N^\prime t} \numhyp^{N^\prime-N^\prime t})$
for solving eq.~\ref{eq:HBA_bellman} with \gls{mcts}. We leave out the dependency on \A{}{\egoidx}, since it affects sample complexities of RSBG and SBG equally.
By increasing \numhyp, we can reduce the sample complexity. Yet, it is mainly dominated by the non-controllable variables and \emph{exponentially depends} on $t$ and $N$ over the sample size of the behavior space $|\BST{}{}{}|$. 

To overcome this problem, we propose a different optimality criterion achieving reduced sample complexity. We combine the optimality criteria of the \gls{rmdp} defined in eq.~\ref{eq:robust_multi_agent} and \gls{sbg} defined in eq.~\ref{eq:HBA_expectation}. We let other agents act adversarially \emph{only within} a hypothesis by defining the worst-case operation over the respective hypothesis action space $\A{\hypidx}{}$. We call this decision model the Robust Stochastic Bayesian Game (RSBG). Specifically, it uses the formal definition of the \gls{sbg}, but with $E^{\a{t}{i}{}}_{\ost{}{}}(\oh{\prime}{})=$
\begin{equation}
\sum_{\hbst{}{-\egoidx}\in \Theta_{-\egoidx}}\text{Pr}(\hbst{}{-\egoidx}|\oh{t}{})\bigg[
\min_{\a{}{-i}{} \in \A{}{-\egoidx}} Q^{\a{}{i,-i}{}}_{\ost{}{}}(\oh{\prime}{})\bigg]
\label{eq:hba_robust}
\end{equation} 
and eq. \ref{eq:HBA_bellman} remaining unchanged. In the next section, we show that with this criterion, we can reduce the sample complexity exponentially compared to the \gls{sbg} for planning over continuous behavior spaces.

\subsection{Monte Carlo Tree Search for the RSBG} \label{sec:MCTS_RSBG}
Planning algorithms incorporating posterior beliefs over types or transition functions are commonly based on variants of \gls{mcts} \cite{guez_efficient_2012,barrett_teamwork_2013}.
We extend the Bayes-adaptive Monte Carlo Planning algorithm \mbox{\cite{guez_efficient_2012}}\footnote{BAMCP converges for discrete action spaces. In continuous action spaces, it may only find a QMDP policy without information gathering behavior \cite{sunberg_online_2017}. We neglect this deficiency since it affects both \gls{sbg} and \gls{rsbg} equally.} to the SBG: At the beginning of each search iteration, we sample a type for each of the other agents $j$ from the posterior belief over types $\hbst{\prime}{j} \sim \text{Pr}(\hbst{\hypidx}{\otheridx}|\oh{t}{})$ and use it in expansion and rollout steps.

To solve the RSBG, we implement the minimum operation in eq.~\ref{eq:hba_robust} over $\A{}{-\egoidx}$ sample-efficiently by letting each other agent \otheridx~ \emph{subjectively} choose a worst-case action at history node $\langle\sh{}\rangle$ within  the hypothesis action space $\A{\hbst{\prime}{j}}{}$.  For this, during back-propagation steps, we maintain expected action-values with respect to agent $i$'s reward function, $Q_j(\langle\sh{}\rangle, \hbst{\prime}{j}, \a{}{}{})$ separately for each hypothesis and other agent $j$. We argue that in tasks with short interaction times a joint action of other agents, consisting of the subjective worst-case actions, is close to their global worst-case action. Such a decoupled action selection results in a sample complexity for the minimum operation equal to the size of only the hypothesis action space $|\A{\hypidx}{}|$. We then get $\BigO(|\HBST{}{-\egoidx}|\cdot|\A{}{-\egoidx}|^t)=\BigO_\text{RSBG}(|\BST{}{}{}|^{t} \numhyp^{N-t})$. The dependency of the sample complexity on the sample size of the behavior space is thus \emph{reduced by factor $N$ in the exponent} compared to $\BigO_\text{SBG}$.

Specifically, we select actions for each agent $j$ with the function \textsc{ActionOtherAgent} in Algorithm~\ref{alg:action_selection_rsbg} called once for each other agent $j\neq i$ in the expansion step.  Line 8 implements the minimum operation of eq.~\ref{eq:hba_robust} returning the worst-case action with respect to agent $\egoidx$ among the set of previously expanded actions $\A{}{j}(\langle\oh{}\rangle, \hbst{\prime}{j})$ from node $\langle\oh{}\rangle$ under type \hbst{\prime}{j}. We propose \emph{hypothesis-based} progressive widening~\cite{couetoux_continuous_2011} in Lines 2-6: Depending on the number of expanded actions $|\A{}{j}(\langle\oh{}\rangle, \hbst{\prime}{j})|$ and the node visit count $N_j(\langle\oh{}\rangle,  \hbst{\prime}{j})$ under hypothesis \hbst{\prime}{j}, we sample a new action from the hypothesis. This approach ensures sufficient exploration of $\A{\hypidx}{}$ to discover the subjective worst case action while guaranteeing a sufficient depth of the search tree.
During roll-out, we only use Line 3 to sample actions for each other agent \otheridx~ according to their currently sampled types $\hbst{\prime}{\otheridx}$.
For the controlled agent $\egoidx$, action selection during expansion and roll-out uses the standard UCB~formula \cite{auer_using_2002}.
\begin{algorithm}[t]
\input{./algorithm_search.tex}
\caption{Adversarial hypothesis-based action selection for MCTS} \label{alg:action_selection_rsbg}
\end{algorithm}

%% file: algorithm_search.tex
\begin{algorithmic}[1]
	\footnotesize 
	\Function{ActionOtherAgent}{$\langle \oh{}\rangle, j, \hbst{\prime}{j}$}
	\If{$|\A{}{j}(\langle\oh{}\rangle, \hbst{\prime}{j})| \leq k_0 N_j(\langle\oh{}\rangle,  \hbst{\prime}{j})^{\alpha_0}$} 
		\State $\a{}{}{j} \gets \p{\hbst{\prime}{j}}(\a{}{\otheridx}{}| \oh{}{})$
		\State append $\a{}{}{j}$ to $\A{}{j}(\langle\oh{}\rangle, \hbst{\prime}{j})$ 
		\State init $Q_j(\langle\oh{}\rangle,  \hbst{\prime}{j}, \a{}{}{j}),\, N_j(\langle\oh{}\rangle,  \hbst{\prime}{j}, \a{}{}{j}) $
		\State \textbf{return} $\a{}{}{j}$
	\Else 
	\State \textbf{return} $\text{arg min}_{\a{}{}{}\in \A{}{j}(\langle\oh{}\rangle, \hbst{\prime}{j})} Q^j(\langle\oh{}\rangle,  \hbst{\prime}{j}, \a{}{}{})  $
	\EndIf
	
	\EndFunction
	
\end{algorithmic}

%% file: experiment.tex
\section{Experiments}

We evaluate the proposed method in two experiments. First, we analyze the performance benefits due to reduced sampling complexity in an artificial crossing domain. For this, we design an artificial crossing environment to have an easy to understand behavior space being closely linked to the intents of the other agents. This allows to assess the performance of the approach for different characteristics of the behavior space with respect to unknown agent intentions.

Then, we use our approach for planning a lane changing task of an autonomous vehicle. Here, we analyze how the approach behaves under more realistic conditions when the unknown behavior of other traffic participants covers a larger multidimensional behavior space.

\subsection{Crossing under Time-Dependent Intents}

\begin{figure}[b]
	\def\svgwidth{0.7\columnwidth}\centering
	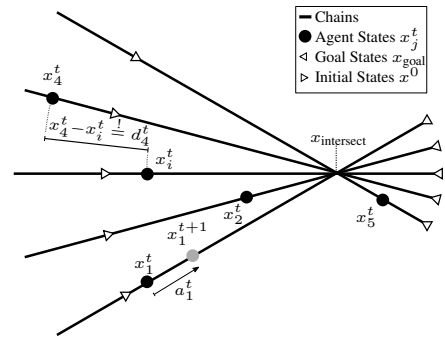 
	\caption{Multi-agent, chain domain with $N=5$ and agent $i=3$.} \label{fig:crossing_environment}
	\vspace{-3mm}
\end{figure}

The crossing domain is depicted in Fig.~\ref{fig:crossing_environment}. Each of the $N=9$ agents moves along its chain with current state $\ost{t}{\otheridx}{=}x_\otheridx^t$ and initial state $\ost{0}{\otheridx}{=}5$. The transition model is  $\ost{t+1}{\otheridx}{=}(x_\otheridx^t {+} \a{t}{j}{})$. State and action space are continuous with $\ost{t}{\otheridx} \in [0, 17],$ and $\a{t}{\otheridx}{}\in [-5, 5]$. The agents' chains intersect at a common point $x_\text{intersect}=15$ which each agent must cross to reach its goal point $x^\otheridx_\text{goal}{=}17 {>} x_\text{intersect}$. Two agents collide when they cross $x_\text{intersect}$ at the same time step $t$.

For the prescribed domain, we define a hypothetical policy $\phyp(\oh{t}{}, \bst{t}{\otheridx}{})$ using a 1-dimensional behavior state $\bst{t}{j}{}{=}\desiredgap{t}{j}$. It models the desired gap \desiredgap{t}{j} of agent \egoidx~ to \otheridx~ with respect to the crossing point: $x_j^t {-} x_\text{intersect} \stackrel{!}{=} x_\egoidx^t{-} x_\text{intersect} {-}\desiredgap{t}{j}$. For positive \desiredgap{t}{j} agent $j$ aims to be behind agent $\egoidx$. For negative \desiredgap{t}{j} agent $j$ aims to be a ahead of agent \egoidx. Parameters $\textsc{Min/MaxVelocity} = -5/{+}5$ define other agent's maximum and minimum action values. The behavioral states of the other agents change randomly over time being sampled from uniform distributions  $\desiredgap{t}{\otheridx} \gets \textsc{sample}(\DistrUniform[\desiredgapleft{\otheridx}, \desiredgapright{\otheridx} ])$ at each time step. In simulation, we want to model that $\ptrue{\otheridx}(\a{t}{j}{}|\oh{t}{}, \ist{t}{j})$, $\ist{t}{j}$ and $\BST{}{\otheridx}{}$ are unknown to agent \egoidx. To avoid the definition of a simulation model based on intention states \ist{t}{}, we apply the hypothetical policy \phyp~ also in simulation: We draw unknown boundaries of behavioral variations $\BST{}{\otheridx}{}=[\desiredgapleft{\otheridx}, \desiredgapright{\otheridx}]$ uniformly from a simulated true behavior space \BST{*}{}{}~($\BST{}{\otheridx}{}\subseteq \BST{*}{}{}$) for each agent and trial. This simulates clearly reasonable intents, such as "give way" ($\desiredgapleft{\otheridx}\gg 0,\, \desiredgapright{\otheridx}\gg 0$) and "take way" ($\desiredgapleft{\otheridx}\ll 0,\, \desiredgapright{\otheridx}\ll 0$), and vague intents changing over time $\ist{t}{j} \neq \ist{t+1}{j}$ ($\desiredgapleft{\otheridx} < 0,\, \desiredgapright{\otheridx}> 0$).

\begin{algorithm}[t]
	\input{./algorithm_behavior.tex}
	\caption{Hypothetical behavior policy for intersection crossing task}
\end{algorithm}

Algorithm 2 gives the implementation of the hypothetical policy $\phyp(\oh{t}{}, \bst{t}{j}{})$ realizing a desired gap \desiredgap{t}{\otheridx}. Line 2 calculates the difference between desired gap and current gap~(\textsc{GapError}) by predicting the position of agent \egoidx~ one time step ahead using its last action. If agent $j$ aims to drive behind agent \egoidx~ ($\desiredgap{t}{\otheridx} {>} 0$), the agent chooses an action exactly the size of the \textsc{GapError} limited by the maximum or minimum velocity. If agent $j$ aims to drive ahead of agent \egoidx~ ($\desiredgap{t}{\otheridx} {<} 0$), the agent additionally avoids to decelerate again, when its last action was larger.

\subsubsection{Planning Algorithms}
Now, we take the role of a domain expert with knowledge of $\phyp(\oh{t}{}, \bst{t}{j}{})$ which must define the full behavioral space \BST{}{}{} by analyzing possible physical situations at the time point of interaction: If agent~\egoidx~ is close to the crossing point ($10{<}\ost{t}{\egoidx}{<}15$), the desired gaps $\desiredgap{t}{\otheridx} \in \BST{}{}{}{=}\{-10, 10\}$ describe all possible behaviors of other agents with respect to agent~\egoidx~ at the time point of interaction. We then use equal-sized partitions of $\BST{}{}{}$ to define the hypothesis set for the RSBG planner following our methodology from sec.~\ref{sec:hypothesis_sets_coverage}. We will study the influence of the parameter \numhyp~ in our experiments.

Based on the MCTS defined in sec.~\ref{sec:MCTS_RSBG} and this hypothesis set, we define the baselines
\begin{itemize}
	\item \textbf{SBG} replacing Line 8 in Algorithm 1 with random  selection among $\A{}{j}(\langle\oh{}\rangle, \hbst{\prime}{j})$,
	\item \textbf{RMDP} using a single hypothesis equivalent to the full behavioral space, $\numhyp=1$ and $\BST{1}{}{}\equiv\BST{}{}{}$,
	\item \textbf{MDP} using a single hypothesis as with RMDP and random action selection as with SBG and
	\item \textbf{SBGFullInfo/RSBGFullInfo} being equal to the SBG, respectively RSBG planners, but having access to the true behavior policies to apply these as hypothesis.
\end{itemize}
Planners RSBG and SBG use the sum posterior defined in~\cite{albrecht_belief_2016} to track the posterior belief over hypothesis. It can deal with zero-probability actions which occur in our hypothesis definition.
\begin{figure*}[t!]
	\centering
	\includegraphics{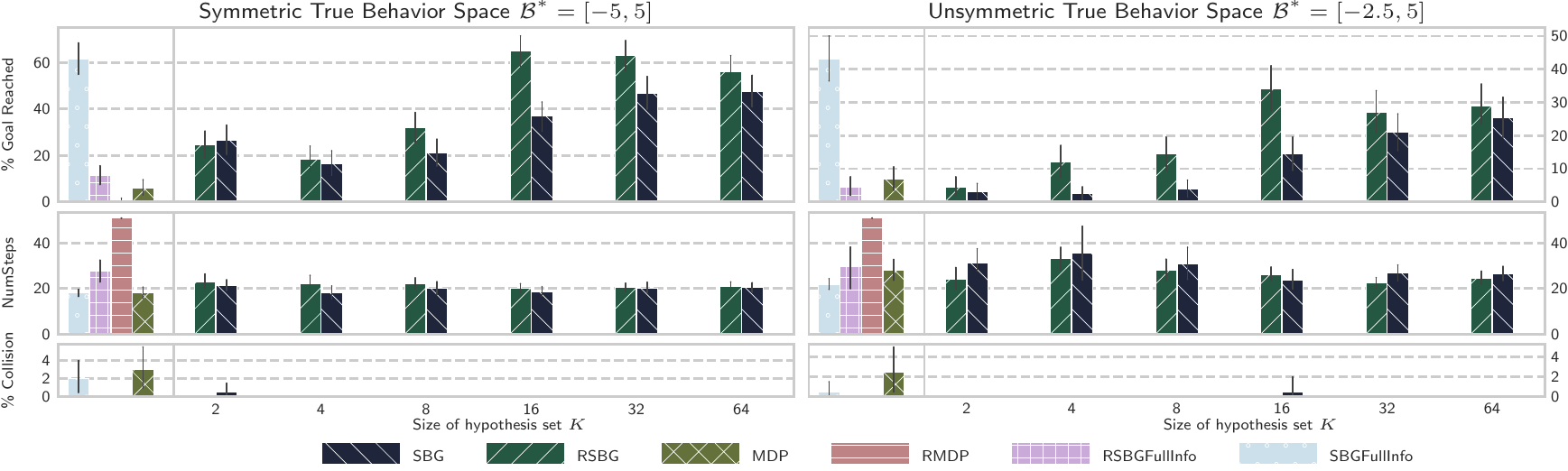}  
	\vspace{-1.5mm}
	\caption{Performance metrics in the crossing domain for RSBG and baseline planners for symmetric and unsymmetric true behavior spaces and, for RSBG and SBG planners for varying size of the hypothesis set \numhyp.  }
	\vspace{-3mm}
	\label{fig:hypothesis_results} 
\end{figure*}
All planners use the reward function $R(\cdot){=}{-}1000\cdot\textsc{collided}{+} 100\cdot\textsc{goal reached}$ and a discrete action space $\a{t}{\egoidx}{} {\in} \A{}{\egoidx} {=} \{-1,0,1,2\}$ for agent~\egoidx, and perform 10000 search iterations in each time step. Progressive widening parameters, $k_0{=}4$ and $\alpha_0{=}0.25$, discount factor $\gamma{=}0.9$ and all other parameters are kept equal for all planners.

\subsubsection{Results}
In our experiment, we simulate the other agents \otheridx~ by sampling a new behavior parameter \bst{t}{\otheridx}{} at every time step from \BST{}{\otheridx}{} and chose their actions with Algorithm 2, respectively. Agent \egoidx~applies one of the planning algorithms to chose an action. Each planner must perform 200 trials. Fixing the random seeds for all sampling operations ensures equal conditions for all planners. We measure the percentage of trials where the agent \egoidx~ reaches the goal, collides or exceeds a maximum number of time steps ($t_\text{max} > 50$). For successful trials, we calculate the average number of time steps to reach the goal.

\begin{figure}[b!]
	\vspace{-3.5mm}
	\centering
	\includegraphics{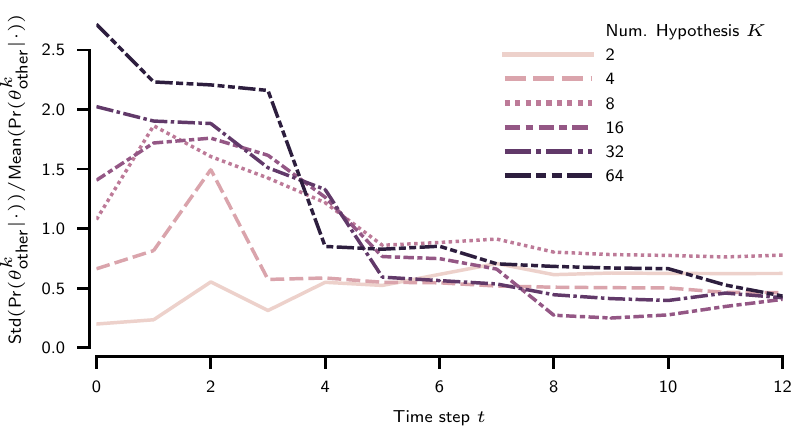}   
	\caption{Normalized standard deviation of the posterior belief.}
	\label{fig:belief_results} 
	
\end{figure}
Fig.~\ref{fig:hypothesis_results} depicts these metrics for the different planners, for SBG and \glspl{rsbg} planners over increasing number of hypothesis \numhyp, and for the case where the true behavior space is symmetric $\BST{*}{}{}=[-5,5]$ (left) and unsymmetric $\BST{*}{}{}=[-2.5,5]$ (right). We leave out the percentage of maximum steps since the percentages sum up to one.
In both settings, the RSBG planner achieves a significantly higher percentage of successful trials for $\numhyp \geq 8$ than the SBG planner. In the case of a symmetric true behavior space, for $\numhyp=16$ and $\numhyp=32$, RSBG achieves equal performance as \mbox{SBGFullInfo} knowing about the true behavior of other agents. The unsymmetric case is more demanding since other agents desire a closer gap to the controlled agent decreasing performance for all planners.
The RSBG and SBG planners achieve nearly equal average of time steps than SBGFullInfo for larger \numhyp. In contrast, the RMDP and RSBGFullInfo planners, purely relying on the worst-case criterion, are overly conservative and mostly exceed the maximum number of allowed time steps. The RSBG planner shows no collisions in contrast to a minor percentage of collisions for the \gls{sbg} planner, and larger percentages resulting with MDP and SBGFullInfo planners. The results  demonstrate that \gls{rsbg}s provide a meaningful compromise between conservative planning with \gls{rmdp}s and riskier planning with \gls{sbg}s and \gls{mdp}s.

We calculate the ratio of sample complexities $\BigO_\text{SBG}/\BigO_\text{RSBG}$ to clarify the advantages of RSBGs for behavior space coverage in our experiment. We set $N=9$, and the prediction time equal to the average number of time steps, $t\approx20$, and get $\BigO_\text{SBG}/\BigO_\text{RSBG} = (|\BST{}{}{}|/K)^{160}$. Defining the required number of samples to cover the full behavior space $|\BST{}{}{}|$ is unclear, but there should be at least one sample for each hypothesis, giving $|\BST{}{}{}|\gg \numhyp$ and thus for fixed \numhyp, $\BigO_\text{SBG}\gg \BigO_\text{RSBG}$. Since both SBG and RSBG planners have the same number of iterations available, RSBG can achieve \emph{better performance due to lower sample complexity for same \numhyp}. It seems that there is an optimal setting of \numhyp=16 for the RSBG planner. For larger \numhyp, the performance of RSBG decreases. Fig.~\ref{fig:belief_results} shows the normalized standard deviation of the posterior belief over agents, hypothesis and ten trials for different \numhyp~ at initial time steps.  With \numhyp=16, the normalized standard deviation stabilizes to the lowest value, indicating a more stable posterior belief. Larger variations in the posterior belief occur at \numhyp=8 or \numhyp>16. We assume that these instabilities counteract a reduction of sample complexity with increasing \numhyp, explaining the observed performance decline for $\numhyp{>}16$, but also the sudden performance increase from  $\numhyp{=}8$ to $\numhyp{=}16$.

Overall, our results indicate that the RSBG decision model performs better than the existing alternatives for planning in continuous behavior spaces. It can plan sample-efficiently at a low number of hypothesis to avoid instabilities in the posterior belief at larger hypothesis sets. 

\subsection{Lane Changing with Multidimensional Behavior Spaces}

Next, we apply the RSBG planner to the problem  of planning a lane change maneuver for an autonomous vehicle in dense traffic where other agents' behavior covers a larger multidimensional behavior space.

\subsubsection{Simulation and Behavior Space Definition}
We use the OpenSource behavior benchmarking environment BARK \cite{bernhard_bark_2020} for simulating the lane changing scenario and controlling the other traffic participants. It is tailored towards a realistic simulation of microscopic traffic scenarios for planning research in autonomous driving.

\begin{figure}[t!]
	\vspace{-3.5mm}
	\centering
	\includegraphics{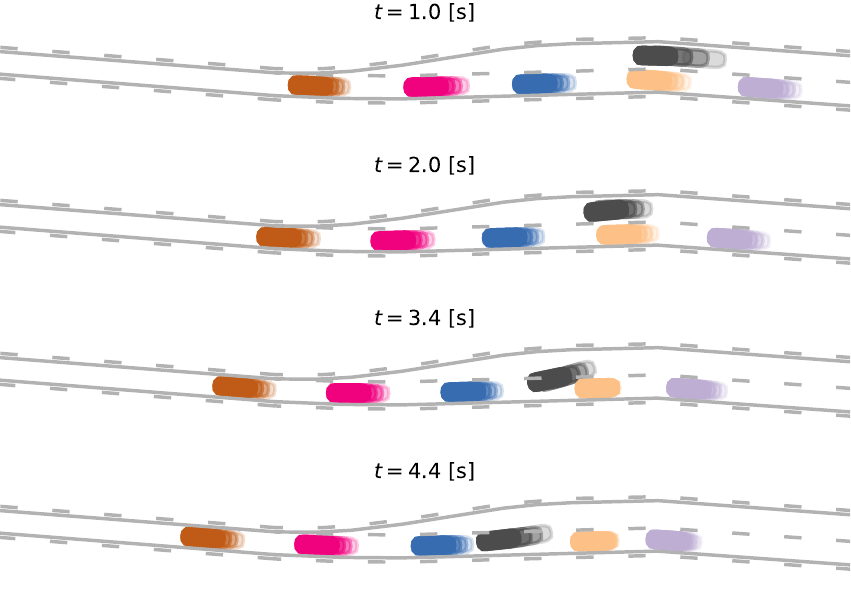}
	\caption{An example of a successful trial of the lane changing task for the RSBG planner at different time points. The agent \egoidx~ (dark grey) must change to the left lane. Past agent positions are indicated with increasing transparency.}
	\label{fig:highway} 
	
\end{figure}

Fig.~\ref{fig:highway} shows a successful trial of the scenario for the RSBG planner. The controlled agent \egoidx~ starts on the right lane (dark grey) and must merge to the left lane where other vehicles are densely placed. We generate different initial starting conditions by sampling the relative distances between vehicles, the position of the controlled agent and the velocities from uniform distributions.

To simulate the behavior of the other vehicles, we use the Adaptive Cruise Control (ACC) model presented by \cite{treiber_traffic_2013}. It combines the Intelligent Driver Model (IDM) \cite{treiber_congested_2000} with a constant acceleration heuristic (CAH). The IDM is a classical car following model which, however, models full emergency braking capability of the following car in any situation. This does not accurately grasp the human nature of driving in more dense traffic where humans also take into account the current accelerations of the leading vehicle. The CAH model assumes that the following vehicle chooses an acceleration close to the leading vehicle. The ACC uses a weighted combination of both models to simulate more human-like following behavior. For mathematical details of the model we refer to \cite{treiber_traffic_2013}. The ACC model defines the following behavior parameters\footnote[2]{The physical acceleration limits are set to -5.0\,$\text{m/}\text{s}^2$ and 8.0\,$\text{m/}\text{s}^2$}.
\begin{itemize}
	\item desired velocity \idmdesiredvelocity~ and desired time headway \idmdesiredheadway~ defining the desired velocity and velocity-dependent safety distance of the driver,
	\item minimum spacing \idmminimumspacing~ defining a minimum distance a driver wants to satisfy,
	\item acceleration factor \idmaccfactor~ describing the acceleration behavior of the driver,
	\item comfortable braking \idmcomftbrake~ defines what acceleration a driver still considers as comfortable,
	\item coolness factor \acccoolness~ is a value between zero and one. A value of one corresponds to the full CAH model. A value of zero to the full IDM model. The higher this value, the more relaxed a driver is in dense situations, avoiding overall harsh braking\footnote[3]{We use $\acccoolness=0.99$ as suggested by \cite{treiber_traffic_2013}.}.
\end{itemize}
Overall, the higher these parameters the more aggressive a driver acts \cite{uhrmacher_multi-agent_2009}. 

Similar to our first experiment, the ACC model defines both the hypothetical \phyp~ and simulated policy \ptrue. For simulation, we define a 5-dimensional true behavior space \BST{*}{5D}{} over these parameters and draw unknown boundaries of behavioral variations $[\bst{l}{}{j, \text{min}}, \bst{l}{}{j,\text{max}}],\, l\in \{1,\ldots,5\}$ for each agent and trial ($\BST{}{\otheridx}{}\subseteq \BST{*}{5D}{}$). We introduce the parameters minimum and maximum boundary widths $\Delta_\text{min}$/$\Delta_\text{max}$ to specify minimum and maximum time-dependent variations of behavior parameters in simulation. This avoids unrealistic large variations of behavior parameters. For building the hypothesis set, we use lower-dimensional behavior spaces to evaluate how variations over multiple behavior parameter dimensions can be captured with hypothesis sets over a smaller set of key behavior parameters. In a preliminary experiment we found that key parameters in the model are the desired safety distance \idmdesiredheadway~ and the desired velocity  \idmdesiredvelocity.  Tab.~\ref{tab:behavior_space} depicts both the simulated and hypothesized behavior space used in our experiment.

\begin{table}[b]
	\tiny
	\input{behavior_space_params}
	\caption{Boundaries  of the simulated true behavior space \BST{*}{5D}{} for the lane changing experiment. We evaluate different full behavior spaces for hypothesis definition defined only over the parameter ranges marked with an X. In the hypothesis definitions, parameters not marked with X are set to the center of the parameter range. } \label{tab:behavior_space}
\end{table}

\subsubsection{Planners}
We benchmark our approach  against the SBG and SBGFullInfo algorithms. The action space of agent \egoidx~ consists of the macro actions lane changing, lane keeping at constant accelerations $\dot{v_\egoidx} {=} \{-5, -1, 0, 1, 4\} [\text{m}/\text{s}^2$] and gap keeping based on the IDM. In this experiment, we fix the number of partitions of each behavior space dimension to 16 which yields $\numhyp_\text{1D} = 16$ and $\numhyp_\text{2D} = 256$. We will evaluate the performance of the algorithms for a low number of iterations being more realistic in real-time critical applications. We apply the same reward function, type of posterior belief update and other parameters as in the previous experiment.

\begin{figure*}[t!]
	\centering
	\includegraphics{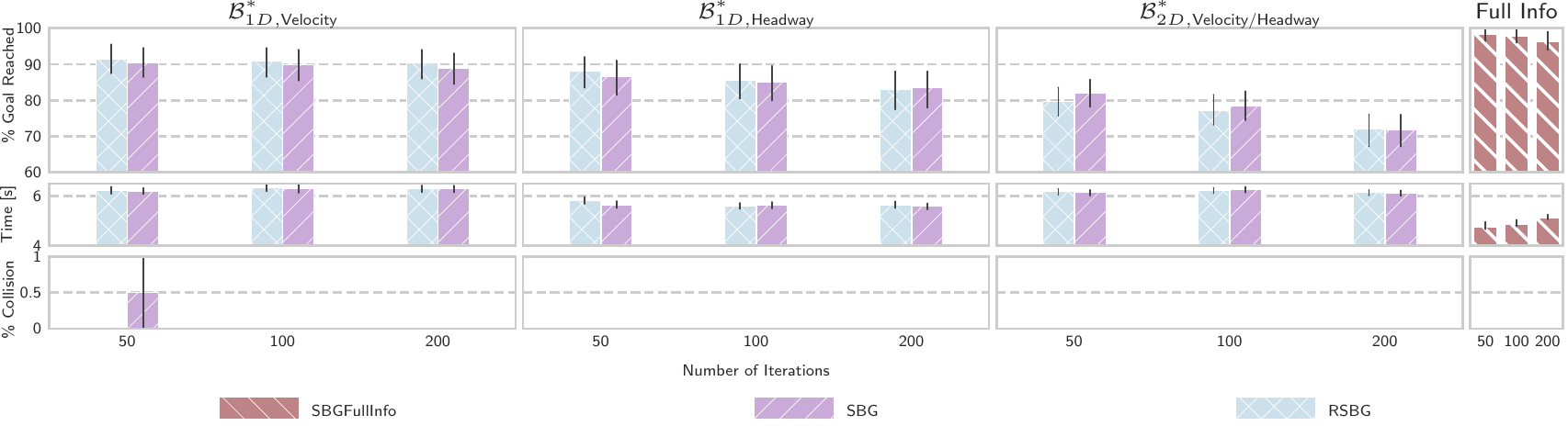}
	\vspace{-1.5mm}
	\caption{Performance metrics in the lane changing task for RSBG, SBG and SGBFullInfo for 1D and 2D full behavior spaces and varying number of search iterations.  }
	\vspace{-3mm}
	\label{fig:hypothesis_results_highway} 
\end{figure*}

\subsubsection{Results}

In our experiment, we simulate the other agents by sampling a new behavior state \bst{t}{\otheridx}{} at every time step from \BST{}{\otheridx}{} and then use the ACC model with this parameters to choose their actions. We perform 200 trials for each planner and evaluate the same criteria as in the previous experiment. The maximum allowed simulation time to solve a scenario is~$t<7.5\,[\text{s}]$.

Fig.~\ref{fig:hypothesis_results_highway} shows the obtained performance metrics when using the 1D or 2D full behavior spaces defined in Tab.~\ref{tab:behavior_space} for different numbers of search iterations.
The RSBG planner marginally outperforms the SBG planner achieving best success rate among all behavior spaces for \BST{}{1D, \text{Velocity}}{}. Both planners achieve a success rate in the region of the SBGFullInfo planner which is remarkable considering that the true behavior space \BST{*}{5D}{} represents a large variation of behavior parameters. This indicates that hypothesis sets over lower dimensional behavior spaces can approximate behavioral variations in larger behavior spaces. The RSBG planner avoids collisions, in contrast to the SBG planner, for 50 search iterations. This suggests that RSBG is more sample-efficient with respect to predicting worst-case outcomes.   

Using \BST{}{1D, \text{Headway}}{}, both planners achieve a lower number of \% goal reached than with \BST{}{1D, \text{Velocity}}{}. However, also the average time to change the lane descreased for both planners with \BST{}{1D, \text{Headway}}{}. This indicates that task requirements have to be taken into account when selecting the behavior parameters of the full behavior space. In our experiment, it seems that it is better to capture the large variations of \idmdesiredvelocity~ via the behavior space  \BST{}{1D, \text{Velocity}}{} to achieve a high success rate. A fast lane change maneuver can however be better achieved when estimating what desired gap other participants aim for. We find that the 2D behavior space does not increase performance compared to using  the 1D behavior spaces. Though, one would expect increased performance due to better capturing of behavior uncertainty, the large number of hypothesis $\numhyp_\text{2D} = 256$ may yield unstable posterior beliefs. As we saw in the previous experiment such instabilities can worsen performance.

Overall, this experiment indicates that our proposed hypothesis design procedure is applicable to practical problems with larger multidimensional behavior spaces. Yet, ways must be found to stabilize the posterior beliefs when using multidimensional full behavior spaces. This may allow a more generic hypothesis design procedure over multiple behavior parameters without sacrificing performance.

%% file: algorithm_behavior.tex
\begin{algorithmic}[1]
	\footnotesize 
    \State \textbf{Output:} $\a{t}{\otheridx}{} = \phyp(\oh{t}{},\bst{t}{j}{})$
    \State $\textsc{GapError} = x_\egoidx^t + \a{t-1}{\egoidx}{} - x_\otheridx^t - \desiredgap{t}{\otheridx}$
	\If{$\desiredgap{t}{\otheridx} > 0$} 
        \If{$\textsc{GapError} < 0$} 
	        \State \textbf{return} $\max(\textsc{GapError}, \textsc{MinVelocity})$
        \Else 
	        \State \textbf{return} $\min(\textsc{GapError}, \textsc{MaxVelocity})$
        \EndIf
    \Else 
	\State \textbf{return} $\max(\min(\textsc{GapError}, \textsc{MaxVelocity}),\a{t-1}{\otheridx}{})$ 
	\EndIf
\end{algorithmic}

%% file: pics/crossing_environment.pdf_tex
\begingroup%
  \makeatletter%
  \providecommand\color[2][]{%
    \errmessage{(Inkscape) Color is used for the text in Inkscape, but the package 'color.sty' is not loaded}%
    \renewcommand\color[2][]{}%
  }%
  \providecommand\transparent[1]{%
    \errmessage{(Inkscape) Transparency is used (non-zero) for the text in Inkscape, but the package 'transparent.sty' is not loaded}%
    \renewcommand\transparent[1]{}%
  }%
  \providecommand\rotatebox[2]{#2}%
  \ifx\svgwidth\undefined%
    \setlength{\unitlength}{425.87229331bp}%
    \ifx\svgscale\undefined%
      \relax%
    \else%
      \setlength{\unitlength}{\unitlength * \real{\svgscale}}%
    \fi%
  \else%
    \setlength{\unitlength}{\svgwidth}%
  \fi%
  \global\let\svgwidth\undefined%
  \global\let\svgscale\undefined%
  \makeatother%
  \begin{picture}(1,0.7316086)%
    \put(0,0){\includegraphics[width=\unitlength,page=1]{crossing_environment.pdf}}%
    \put(0.76911743,0.43128862){\color[rgb]{0,0,0}\makebox(0,0)[b]{\smash{\tiny $x_\text{intersect}$}}}%
    \put(0.33943734,0.15023278){\color[rgb]{0,0,0}\makebox(0,0)[b]{\smash{\tiny $x_1^t$}}}%
    \put(0.83078864,0.25398043){\color[rgb]{0,0,0}\makebox(0,0)[b]{\smash{\tiny $x_5^t$}}}%
    \put(0.53678454,0.2606219){\color[rgb]{0,0,0}\makebox(0,0)[b]{\smash{\tiny $x_2^t$}}}%
    \put(0.13627826,0.5687481){\color[rgb]{0,0,0}\makebox(0,0)[b]{\smash{\tiny $x_4^t$}}}%
    \put(0.36254043,0.38151566){\color[rgb]{0,0,0}\makebox(0,0)[lb]{\smash{\tiny $x_\egoidx^t$}}}%
    \put(0,0){\includegraphics[width=\unitlength,page=2]{crossing_environment.pdf}}%
    \put(0.43517398,0.21497352){\color[rgb]{0,0,0}\makebox(0,0)[b]{\smash{\tiny $x_1^{t+1}$}}}%
    \put(0.42855619,0.09167122){\color[rgb]{0,0,0}\makebox(0,0)[b]{\smash{\tiny $\a{t}{1}{}$}}}%
    \put(0,0){\includegraphics[width=\unitlength,page=3]{crossing_environment.pdf}}%
    \put(0.23971944,0.44553045){\color[rgb]{0,0,0}\rotatebox{-6.43234713}{\makebox(0,0)[b]{\smash{\tiny $x_4^t {-} x_\egoidx^t \stackrel{!}{{=}}\desiredgap{t}{4}$}}}}%
    \put(0.71652053,0.60843686){\color[rgb]{0,0,0}\makebox(0,0)[lb]{\smash{\tiny Goal States $x_\text{goal}$}}}%
    \put(0,0){\includegraphics[width=\unitlength,page=4]{crossing_environment.pdf}}%
    \put(0.71652053,0.56235016){\color[rgb]{0,0,0}\makebox(0,0)[lb]{\smash{\tiny Initial States $x^{0}$}}}%
    \put(0,0){\includegraphics[width=\unitlength,page=5]{crossing_environment.pdf}}%
    \put(0.71652053,0.69598804){\color[rgb]{0,0,0}\makebox(0,0)[lb]{\smash{\tiny Chains}}}%
    \put(0,0){\includegraphics[width=\unitlength,page=6]{crossing_environment.pdf}}%
    \put(0.71652053,0.65120719){\color[rgb]{0,0,0}\makebox(0,0)[lb]{\smash{\tiny Agent States $x_\otheridx^t$}}}%
  \end{picture}%
\endgroup%

%% file: behavior_space_params.tex
\begin{tabular}{p{0.15\columnwidth}*{2}{>{\centering\arraybackslash}p{0.15\columnwidth}}*{3}{>{\centering\arraybackslash}p{0.09\columnwidth}}} 
    \toprule
    &  \multicolumn{2}{c@{}}{\BST{*}{5D}{} } & \BST{}{1D, \text{Vel.}}{} & \BST{}{1D, \text{Head.}}{} & \BST{}{2D}{} \\
    \midrule 
    Param \bst{l}{}{} & [\bst{l}{}{\text{min}},\bst{l}{}{\text{max}}] & $\Delta_\text{min}$/$\Delta_\text{max}$ &  &  &  \\ 
    \midrule 
   \idmdesiredvelocity\, [m/s] 		& [5.0, 15.0] & 5.0 / 10.0   &  X &   &   \\
   \idmdesiredheadway\, [s] 		& [0.0, 1.0] & 0.5 / 1.0   &   & X & X \\
   \idmminimumspacing\, [1]   				& [0.0, 0.5] & 0.0 / 0.5   &   &   &   \\ 
   \idmaccfactor\, [m/$\text{s}^2$] & [1.0, 2.0] & 0.5 / 1.0  &   &   &   \\
  \idmcomftbrake\, [m/$\text{s}^2$] & [2.0, 3.0] & 0.8 / 1.0  &   &   &   \\ 
    \bottomrule
    \end{tabular}

%% file: conclusion.tex
\section{Conclusion}

This work proposes a novel prediction model for self-interested agents in multi agent systems based on physically interpretable behavior spaces, and an accompanying hypothesis design process ensuring that a set of behavior hypothesis covers all physically realistic behavioral variations. We propose a novel decision-theoretic framework under this paradigm, the \gls{rsbg}, combining \gls{rmdp}s~\cite{nilim_robust_2005-1} and SBGs~\cite{albrecht_game-theoretic_2013}, and theoretically identify that, compared to SBGs, the sample complexity of RSBGs for planning with MCTS is exponentially reduced under our behavior space model. In an intersection crossing task, we empirically demonstrate that the RSBG planner outperforms the state-of-the-art planners by a large margin, achieving the same performance as a planner knowing of other agents' true behavior. In a lane changing task, we show that our proposed hypothesis design procedure is applicable to practical problems with larger multidimensional behavior spaces. 

In future, we plan to improve hypothesis definitions over larger behavior spaces by finding ways to stabilize belief tracking for multidimensional behavior spaces. To assess the performance of our approach under practical conditions, we plan to extract behavior spaces from recorded driving data.